\documentstyle[12pt]{article}

\begin{document}

\begin{center}
{\bf AMPLIFICATION EFFECTS ON THE TRANSMISSION AND REFLEXION
PHASES IN 1D PERIODIC SYSTEMS}\\
\vspace{1cm}

Nouredine Zekri \footnote{Corresponding
author, e-mail: nzekri@meloo.com}, Karima Bennabi and Souad Maarouf  \bigskip

{\em U.S.T.O., D\'{e}partement de Physique, L.E.P.M.,\\
B.P.1505 El M'Naouar, Oran, Algeria}\bigskip .{\em \\
and }\bigskip

{\em Abdus Salam International Centre for Theoretical Physics, Trieste
(Italy)}.
\end{center}

\baselineskip=24pt

\centerline{\bf Abstract} \bigskip

\hspace{0.33in}We investigate the localization observed recently 
for locally non-hermitian Hamiltonians by studying the effect of the amplification
on
the scaling behavior of the transmission and reflection phases  
in 1D periodic chains of $\delta$-potentials. The amplification here is
represented by an imaginary term added to the on-site potential. It is found
that both phases
of the transmission and reflection amplitudes are strongly affected by the 
amplification term.  In particular, the phases in the region of amplification 
become independent of the length scale while they oscillate strongly near the 
maximum transmission (or reflection). The interference effects on the phase in 
passive systems are used to interpret those observed in the presence of
amplification. The phases of the transmission and reflection are found to 
oscillate in passive systems whith increasing periods in
the allowed band for the transmission phase while for the reflection phase, its 
initial value is always less than $\pi /2$ in this band.

\vspace{2cm} \noindent {Keywords:} Non hermiticity, Localization,
Amplification. \vspace{1cm}


\newpage

\section{Introduction }

\hspace{0.33in} 
Recently, there was an increase of interest in non-hermitian hamiltonians and
quantum phase transitions (typically localized to extended
wavefunctions) in systems characterized by 
them. There are in general two classes of problems in this context: one in which the 
non-hermiticity is in the nonlocal part \cite{nlnh1,nlnh2} and the other in which it is in the local 
part [3-8]. In the first category, one considers an imaginary vector potential added to the 
momentum operator in the Schr\"odinger hamiltonian. 
In the second category (non-hermiticity in the local term), an imaginary term is introduced in 
the one-body potential. It is well-known from textbooks on quantum mechanics that 
depending on the sign of the imaginary term, this means the presence of a sink (absorber) 
or a source (amplifier) in the system. It may be noted that this second category does also 
have a counterpart in classical systems characterized by a Helmholtz (scalar) wave equation 
as well, where the practical application is in the studies of the effects of classical wave 
(light) localization due to backscattering in the presence of an amplifying (lasing) medium 
that has a complex dielectric constant with spatial disorder in its real part 
\cite{pradhan,zhao}. There is a common thread binding both the problems though, namely 
that the spectrum for both becomes complex (the hamiltonian being non-hermitean or real 
non-symmetric), but can admit real eigenvalues as well. The common property is that the 
real eigenvalues represent localized states and the eigenvalues off the real lines extended 
states. That it is so in the first category has been shown in the recent works starting with 
Hatano and Nelson and followed by others \cite{nlnh1,nlnh2}. In the rest of the
paper we would be concerned with non-hermitian hamiltonians of the second category 
only. For this category with sources at each scatterer and in the absence of impurities, 
it seems counter-intuitive that there are localized solutions; but it has been shown in 
a simple way [5,8,9] that the real eigenvalues
are always localized. 
However, up to now the physical origins of this effect have not been provided.
Since the localization is a consequence of the backscattering and the destructive
interferences, we expect this effect to be related to the scaling behavior of the
phases of the transmission and reflection amplitudes. This is the aim of this
letter where we examine numerically the effect of the
amplification on the phase of the transmission and reflection amplitudes. We
use for this end the Kronig-Penney model which is a continuous multiband model. We 
first consider a periodic passive system in 
order to understand the behavior of the phase for localized and extended
states. This allow us to explain the phase behavior in
such amplifying systems.  

\section{Model description}

\hspace{0.33in} We consider a non interacting electron of energy $E$
moving through a linear chain of $\delta $-potentials strengths
strength $\beta_{n}$, $n$ is the site position. In each site an imaginary
term $\eta$ is included leading to a Non Hermitian Hamiltonian.
The Schr\"{o}dinger equation then reads

\begin{equation}
\left\{ -\frac{d^{2}}{dx^{2}}+\sum_{n}(\beta _{n}+\eta)\delta (x-n)\right\}
\Psi (x)=E\Psi (x)
\end{equation}

\noindent Here $\Psi (x)$ is the single particle wavefunction at $x$,
and $E$ is expressed in units of $\hbar ^{2}/2m$ with
$m$ being the free electron effective mass. For simplicity, the lattice
spacing is taken to be unity in all this work. Since we are interested only
in periodic systems, the potential strength $\beta_{n}$ is a constant
$\beta_{0}$. The complexe potential appearing in the local part of the Hamiltonian
in (1) leads either to complex eigenvalues and real wavenumbers or real eigenvalues
and complex vavenumbers. We consider the system Ohmically connected to ideal
leads so that the second case is used since the total energy is conserved.
In this case the imaginary part acts either
as a sink (absorber) if $\eta<0$ or as a source (amplifier) if $\eta>0$ \cite{zekri1}.
From the computational point of view it is more useful to
consider the discrete version of the Schr\"{o}dinger equation which is called
the generalized Poincar\'{e} map and can be derived without any approximation
from (1). It reads \cite{Bellis}

\begin{equation}
\Psi _{n+1} = \left[ 2\cos k + \frac{\sin k}{k}( \beta_0 + i \eta
) \right] \Psi _{n} - \Psi _{n-1}
\end{equation}

\noindent where $\Psi _{n}$ is the value of the wavefunction at site $n$ and 
$k=\sqrt{E}$. This representation relates the values of the wavefunction at
three successive discrete locations along the x-axis without restriction on
the potential shape at those points and is very suitable for numerical
computations. The solution of equation (2) is done iteratively by taking for
our initial conditions the following values at sites $1$ and $2$ : $\Psi
_{1}=$ $\exp (-ik)$ and $\Psi _{2}=$ $\exp (-2ik)$. We consider here an
electron having a wave number $k_{F}$ (at Fermi energy) incident at site $%
N+3 $ from the right (by taking the chain length $L=N$, i.e. $N+1$
scatterers). The transmission and reflection amplitudes ($t$ and $r$) can
then be expressed as

\begin{equation}
t=\frac{-2i\exp (-ik(N+3))\sin k}{\Psi _{N+3}\exp (-ik)-\Psi _{N+2}},
\end{equation}

\noindent and

\begin{equation}
r=\frac{\exp (-2ik(N+3))\left( \Psi _{N+2}-\exp (ik)\Psi _{N+3}\right) }{%
\Psi _{N+3}\exp (-ik)-\Psi _{N+2}},
\end{equation}

\noindent where the terms $\exp (-ik(N+3))$ and $\exp (-2ik(N+3))$
apprearing respectively in the transmission and reflection amplitudes
originate from the fact that the electron is incident at site $N+3$ with an
incident phase $-k(N+3)$. Therefore, these fictious phases are to be
disgarded. Note here that the wave number $k$ appearing in the last expressions
is that of the free electron moving in the leads and is different from that inside
the system (which is complex). From Eqs. (3 and 4) the phases of the transmission
and reflection
amplitudes depend only on the values of the wavefunction at the end sites, $%
\Psi _{N+2}$, $\Psi _{N+3}$ which are evaluated from the iterative equation
(2). The phases of the transmission and reflection amplitudes ($\Phi_t$ and $\Phi_r$)
are then the arguments of $t$ and $r$ respectively. These phases vary
obviously between $0$ and $2 \pi$. 

\section{Results and discussion}

\hspace{0.33in} As discussed below, the observed asymptotic localization in
amplifying periodic systems \cite{zekri1} should come from the phase
interferences and the backscattering. Indeed, the maximum transmission length
($L_{max}$) in this case can be seen as the characteristic length separating the
region where the amplification dominates from that where the interferences and
backscattering dominate (below $L_{max}$). Let us first consider the effect on the
transmission and reflection phases. 

\hspace{0.33in} In order to understand the phase behavior in the case of
constructive and destructive interferences, we start examining its scaling in a
passive periodic system. We fix in this case $\beta_0 =8$ which, from Eqs. (1)
and (2) leads us to the energy spectrum shown in Fig.1. In this spectrum, we
choose the energies $E=1$, $E=3$ and $E=5$ to scan the
phase scaling either in the gap and the allowed band (Figs.2). The
transmission phase in Fig.2a oscillates around $\pi$ with decreasing periods for
energies away from the allowed band while they increase inside this band.
Therefore a higher frequency oscillating phase means a localization. In Fig.2b,
the initial reflection phase seems to be always between $\pi /2$ and $3 \pi
/2$ for energies in the gap which corresponds to localized states for such
finite systems. 

\hspace{0.33in} Let us now examine the phase scaling for amplifying systems 
$\eta >0$ (see Figs.3). For simplicity we consider that the on-site potential is
purely imaginary (i.e., $\beta =0$). We see in particular in these figures that both
the reflection and amplification phases remain constant in the region where the
transmission coefficient grows. It is important to notice that the reflection phase
is greater than $\pi /2$ which indicates that there are destructive interferences 
in the region of growing transmission but they seem to not affect it. In the region
of maximum transmission (and reflection) both phases oscillate and the transport
properties of the system seems to become sensitive to them.

\section{Conclusion}

\hspace{0.33in} We used in this letter the effect of the amplification on
the scaling behavior of both transmission and reflection phases in order to
interpret the recently observed effect on the coefficients. The main results
show a constant phase in the growth region while it starts oscillating near the
maximum transmission and reflection. However, the amplification effect has been
studied here only in the allowed band of the corresponding passive periodic system
(since $\beta_0 = 0$ when the amplification $\eta$ is applied, all the spectrum
of the passive system is Bloch like). Therefore, it is interesting to examine this
effect in the gap of the corresponding passive system. In this case the transmission
coefficient is exponentially decaying (the system being finite) and the Lyapunov
exponent should be affected differently by the amplification.
 
\vspace{0.2in}

\noindent{\bf Acknowledgements} NZ would like to thank the ICTP for its
hospitality and the Arab Funds for its support during the progress of this
work.

{\bf \newpage }

\begin{center}
{\bf Figure Captions }
\end{center}

\bigskip

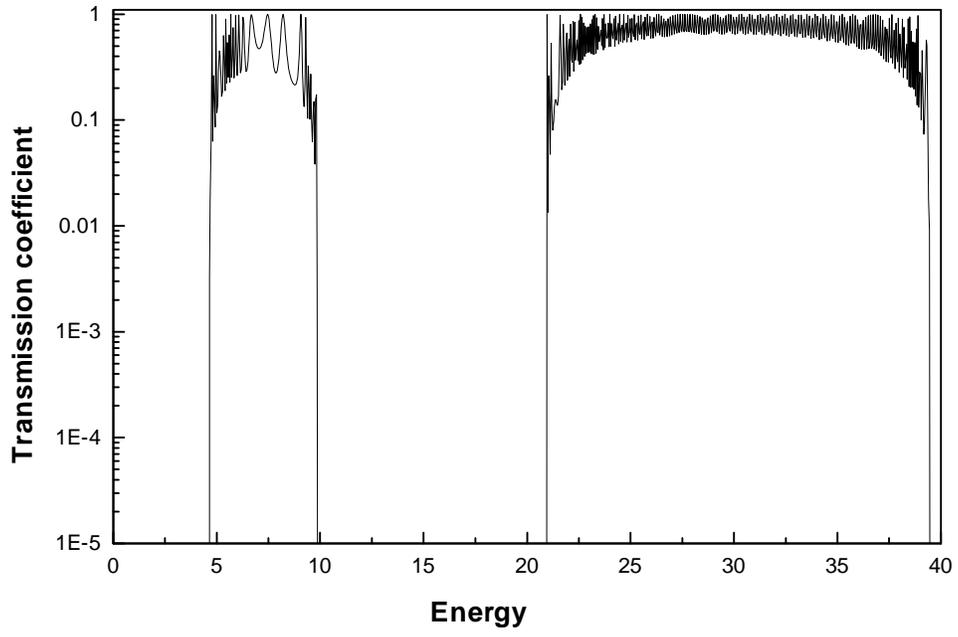
{\bf Fig.1} Transmission coefficient (in a log scale) versus energy for
$\beta_0 = 8$ and $\eta =0$ (passive system).

\bigskip
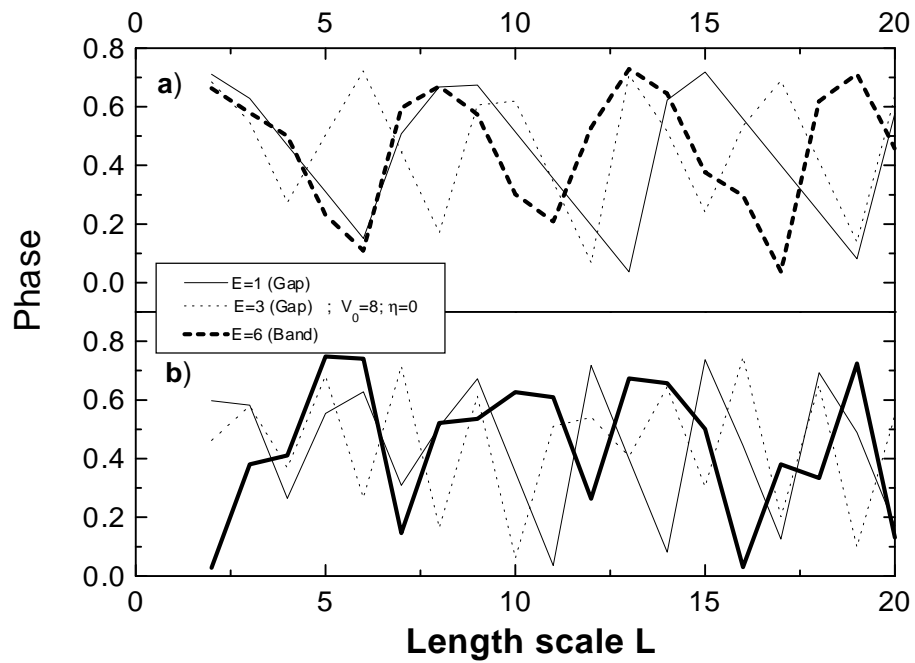
{\bf Fig.2} Variation of the reflection and transmission phase with the length scale
for $\eta=0$, $\beta_0 = 8$ and different energies $1, 3$ and $5$. a) $\Phi_t$, b)
$Phi_r$

 \bigskip
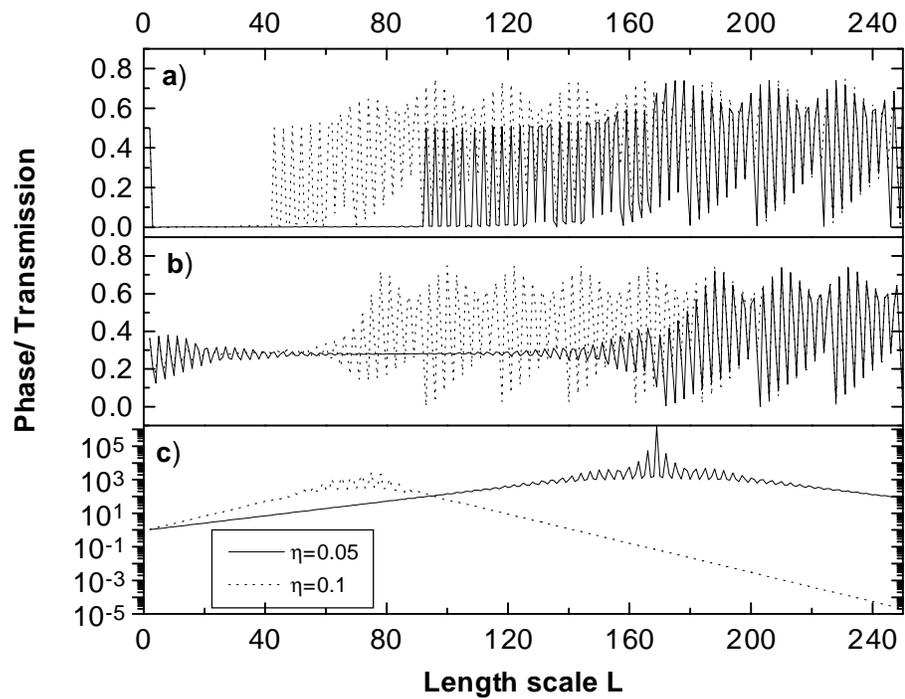
{\bf Fig.3} variations of the reflexion and transmission phases and the
transmission coefficient with the
length scale $L$ for $\beta =0$, $\eta =0.05$ and $0.1$ and the energy $E=1$. a)
phase of the transmission, b) phase of the reflection, c) transmission coefficient.

\end{document}